\documentclass[10pt,twocolumn,aps,showpacs,prc]{revtex4-2}
\usepackage[T1]{fontenc}
\usepackage{graphicx}
\usepackage{amsmath,bm}
\usepackage{amssymb}
\usepackage{xcolor}
\usepackage{hyperref}
\begin{document}

\title{Properties of glitching pulsars in the Skyrme-Hartree-Fock framework}

\author{Joydev Lahiri$^{1\dag}$, Debasis Atta$^{2\S}$ and D. N. Basu$^{3\dag}$}

\affiliation{$^\dag$Variable Energy Cyclotron Centre, 1/AF Bidhan Nagar, Kolkata 700064, India}
\affiliation{$^\S$Government General Degree College Kharagpur-II, West Bengal 721149, India}

\email[E-mail 1: ]{joy@vecc.gov.in}
\email[E-mail 2: ]{debasisa906@gmail.com}
\email[E-mail 3: ]{dnb@vecc.gov.in}

\date{\today }

\begin{abstract}
	
    We address the issues of crustal properties of neutron stars such as crustal mass, crustal radius, crustal fraction of moment of inertia and investigate the crustal and structural properties related to the glitching mechanism observed in pulsars. The mass, radius and crustal fraction of moment of inertia in neutron stars have been determined using $\beta$-equilibrated (npe$\mu$) dense neutron star matter obtained using the extended Skyrme effective interactions with NRAPR and Brussels-Montreal parameter sets. The maximum mass of neutron star calculated from these sets is able to reach $\sim$2$M_\odot$ and higher, corroborating the recently observed masses of compact stars. The crustal fraction of the moment of inertia depends sensitively on the pressure and corresponding density at core-crust transition. The core-crust transition density and pressure together with the extracted minimum crustal fraction of the total moment of inertia provide a limit for the radii of pulsars. Present calculations imply that due to crustal entrainment the crustal fraction of the total moment of inertia is about 5.5$\%$.		

\vskip 0.2cm
\noindent

\end{abstract}
\maketitle

\noindent
\section{Introduction}
\label{section1}

    After the path breaking works of Baade-Zwicky \cite{Ba34} and Oppenheimer-Volkoff \cite{Op39} followed by the identification of pulsars as rotating neutron stars (NSs) about half a century ago \cite{Go68}, several studies were made to resolve the conundrum of NS structure. It is now well recognized that the magnitude of variation of density from the surface to core is of about fifteen orders. In order to understand such complex structure, precise knowledge of the neutron star matter equation of state (EoS) in the different density regions is required. The regions of very low density, the sub-nuclear density and from neutron drip density to about nuclear density can be well described by Feynman-Metropolis-Teller (FMT) \cite{FMT}, Baym-Pethick-Sutherland (BPS) \cite{Ba71} and Baym-Bethe-Pethick (BBP) \cite{Ba71a} EoSs, respectively. In the present work NS structure is studied using a composite EoS, i.e. FMT, BPS, BBP and the EoS of the $\beta$-equilibrated dense neutron star matter with progressively increasing densities. The nuclear matter EoS in the high density domain suffers from uncertainties. A reliable EoS should have consistency in describing the nuclear equilibrium properties \cite{Bl80} near ground state and the hot and dense nuclear matter created in high energy collisions \cite{St86} as well. It provides inputs in hydrodynamic simulations which try to find out whether a star at the later stage of its life explodes or not \cite{Br88} and how a NS is created. Moreover, the EoS has a bearing on whether or not the direct URCA (dURCA) process, whereby a NS cools very rapidly \cite{La91,Pe92}, can take place in NSs. 
    
    A nuclear matter EoS valid under extreme conditions is an essential tool that incorporates features of the nuclear force and for applications to astrophysics requires its precise knowledge for a wide range of densities (from subsaturation to suprasaturation) having a wide range of isospin asymmetries. In order to test our understanding on nuclear matter under extreme conditions, a natural laboratory is provided by NSs which are among the most dense objects in the universe. Both for the astrophysics and the nuclear physics communities, understanding their structure and properties has long been a very challenging task \cite{La04}. 
   
    It is an accepted fact that the properties of the crust play an important role in interpreting many astrophysical observations \cite{Ba71,Ba71a,Pe95,Pe95a,La01,La07,St05,Li99,Ho04,Ho04a,Bu06,Ow05,Ru06}. The inner crust spans the region from the neutron drip point to the inner edge separating the solid crust from the homogeneous liquid core. While the neutron drip density $\rho_d$ is relatively well determined \cite{Ba71}, the transition density $\rho_t$ at the inner edge is rather uncertain particularly because of the fact that knowledge on EoS is somewhat limited, especially the density dependence of the symmetry energy of nuclear matter \cite{La01,La07}. At the inner edge of the crust a phase transition occurs from the inhomogeneous matter at lower densities to the high-density homogeneous matter. When the uniform neutron-proton-electron-muon (npe$\mu$) matter becomes unstable with respect to the separation into two coexisting phases (one corresponding to nuclei, the other to a nucleonic sea) \cite{La07}, the transition density attains its critical value $\rho_t$.

    The calculation of the transition density $\rho_t$ is a very complex issue as the inner crust may have a very complicated structure. The canonical approach is to search for the density at the point where the uniform fluid starts to become unstable against small-amplitude density fluctuations, signaling the creation of the nuclear clusters. This formalism includes the dynamical \cite{Ba71,Ba71a,Pe95,Pe95a,Do00,Oy07,Du07,Xu09,Ts19,Ma21}, the thermodynamical \cite{La07,Ma21,Ku04,Ku07,Wo08,Mo12} and the random phase approximation (RPA) \cite{Ho01,Ca03} approaches. The different density regions of a compact star are regulated by different EoSs. The density domain can be largely classified into two distinct regions: a crust which is responsible for $\sim$ 0.5$\%$ of mass and $\sim$ 10$\%$ of the radius of a star. The core accounts for the rest of the mass and radius of the star. Except in the outer few meters, the outer layers consist of a solid crust about a km thick comprising a lattice of bare nuclei immersed in a degenerate electron gas. When one penetrates into the crust deeper, the nuclear species become progressively more neutron rich because of the rising electron Fermi energy, beginning in principle from $^{56}$Fe through $^{118}$Kr at mass density $\approx$ 4.3 $\times$ 10$^{11}$ g cm$^{-3}$ \cite{Ba75}. This density corresponds to the neutron drip point where the nuclei are so much neutron rich that with any further increase in density, the continuum neutron states start filling up. Consequently, the neutron-rich nuclear lattice gets permeated by a sea of neutrons.

    In this work we use EoSs which cover the crustal region of a compact star to be FMT, BPS, BBP and low density part of neutron matter up to core-crust transition density. The $^{56}$Fe nucleus formed at the endpoint of thermonuclear burning is the most energetically favorable one at the low densities. The FMT \cite{FMT} is based on Thomas-Fermi model. The outermost crust is essentially iron at high pressure together with a fraction of the electrons bound to the nuclei. The main problem in the derivation of this EoS was the calculation of the electronic energy. The EoS of BPS \cite{Ba71} is appropriate at subnuclear densities ranging from $\sim$ 10$^4$ g cm$^{-3}$ to 4.3 $\times$ 10$^{11}$ g cm$^{-3}$, the neutron drip density. The EoS of BPS incorporates the effects of the lattice Coulomb energy on the equilibrium nuclide. The EoS of BBP \cite{Ba71a} is applicable in the region from neutron drip density to about nuclear density 2.5 $\times$ 10$^{14}$ g cm$^{-3}$. This domain is composed of nuclei, electrons and free neutrons. The derivation of EoS is based on a compressible liquid drop model with conditions that the free neutron gas must be in equilibrium with neutrons in nuclei which must be stable against $\beta$-decay.
    
    In the present work, we investigate the structural and crustal properties of neutron stars, such as mass-radius relationship, crustal mass ($\Delta M$), crustal radius ($\Delta R$) and crustal fraction of moment of inertia ($\Delta I/I$) in order to analyze those for glitching pulsars using the EoS for neutron-rich nuclear matter obtained from the NRAPR set of Skyrme NN-interaction parameters \cite{St05}.
     The $\beta$-equilibrium in dense npe$\mu$ matter has been explored. The equilibrium proton fraction has been calculated and possibility of dURCA process has been studied. The mass-radius relationship
of NSs as well as the maximum mass allowed for NS have been determined and crustal properties have been probed using the present interaction. 

\noindent
\section{Theoretical formalism}
\label{section2}

    T. H. R. Skyrme developed the Skyrme interaction in 1959 \cite{Sk56}. Later, Vautherin and Brink performed Hartree-Fock (HF) calculations for spherical nuclei using Skyrme's density-dependent effective NN interaction. They found a remarkable description of doubly-closed shell nuclei ground-state properties \cite{Va72}. Since Skyrme's pioneering work, and the Brink and Vautherin parametrization of the original interaction, constant attempts have been made by various groups to modify the Skyrme-type effective NN interaction parameters and confine Skyrme interaction parameter sets in order to better recreate experimental results. The analytical simplicity of the interactions is a main advantage allows one to determine the parameters that include fundamental properties.

\subsection{Skyrme interaction and equation of state}  
    
    The standard form of the Skyrme interaction is given by \cite{Ch97} 
\begin{eqnarray}
V(\bm{r}_1,\bm{r}_2)&=&t_0(1+x_0P_{\sigma})\delta(\bm{r})  \notag \\
& &+\frac{1}{2}t_1(1+x_1P_{\sigma})[\bm{k}'^2\delta(\bm{r})+\mathrm{c.c.}] \notag \\
& &+t_2(1+x_2P_{\sigma})\bm{k}'\cdot\delta(\bm{r})\bm{k} \notag \\
& &+\frac{1}{6}t_3(1+x_3P_{\sigma})\rho ^{\alpha}\left(\bm{R}\right)\delta(\bm{r})\notag\\
& &+iW_0(\bm{\sigma}_1+\bm{\sigma}_2)\cdot[\bm{k}'\times\delta(\bm{r})\bm{k}],
\label{Eq:Sky}
\end{eqnarray}
\noindent
where $\bm{r}=\bm{r}_1-\bm{r}_2$, $\bm{R}=(\bm{r}_1+\bm{r}_2)/2$, $\bm{\sigma}_i$ is the Pauli spin operator, $P_{\sigma}$
is the spin-exchange operator,
$\bm{k}=-i(\bm{ \nabla}_1-\bm{\nabla}_2)/2$ is the relative
momentum operator, and
$\bm{k}^{\prime}$ is the conjugate operator of $\bm{k}$ acting
on the left. In the above equation the first term represents central term, second and third terms represent non-local term, fourth term represents density dependent term and last term represents spin-orbit term. The above Skyrme interaction may further be generalized by the addition of zero-range density- and momentum-dependent terms~\cite{Kre77,Ge86,Zhu88,Cha09,Gor10,Gor13,Gor15}

\begin{eqnarray}
\frac{1}{2}t_4(1+x_4P_{\sigma}) 
[\bm{k}'^2\rho^{\beta}\left(\bm{R}\right)\delta(\bm{r})+\mathrm{c.c.}] \notag\\
+t_5(1+x_5P_{\sigma})\bm{k}'\cdot\rho^{\gamma}\left(\bm{R}\right)\delta(\bm{r})\bm{k},
\label{Eq:ExSky}
\end{eqnarray}
\noindent
These are the density dependent generalization of the $t_1$ and
$t_2$ terms in Eq.~(\ref{Eq:Sky}). Infinite nuclear matter being spatially homogeneous, the axis can not be defined which leads to the absence of spin-orbit coupling. This implies that $W_0$ term does not contribute in nuclear matter calculations. Thus the generalized Skyrme interaction collectively requires 15 parameters, {\it viz.} $t_0$, $t_1$, $t_2$, $t_3$, $t_4$, $t_5$, $x_0$, $x_1$, $x_2$, $x_3$, $x_4$, $x_5$, $\alpha$, $\beta$ and $\gamma$. As given in Table~\ref{table1}, different values of these parameters generate the NRAPR \cite{St05} and the Brussels-Montreal \cite{Gor13} interactions used in this paper. Using density functional theory the energy per particle of an asymmetric infinite nuclear matter (ANM) may be expressed as \cite{Du12}

\begin{multline}\label{seqn2}
\epsilon(\rho,\eta) = 
\frac{3\hbar^2}{20}\left[\frac{1}{M_\mathrm{n}}(1+\eta)^{5/3} +
\frac{1}{M_\mathrm{p}}(1-\eta)^{5/3}\right]nk_\mathrm{F}^2    \\
 + \frac{1}{8}t_0\Biggl[3 - (1+2x_0)\eta^2\Biggr]n^2  \\
 + \frac{3}{40}t_1\Biggl[(2+x_1)F_{5/3}(\eta) -
\left(\frac{1}{2}+x_1\right)F_{8/3}(\eta) \Biggr]n^2k_\mathrm{F}^2 \\
 + \frac{3}{40}t_2\Biggl[(2+x_2)F_{5/3}(\eta) +
\left(\frac{1}{2}+x_2\right)F_{8/3}(\eta) \Biggr]n^2k_\mathrm{F}^2 \\
 + \frac{1}{48}t_3\Biggl[3-(1+2x_3)\eta^2\Biggr]n^{\alpha+2} \\
 + \frac{3}{40}t_4\Biggl[(2+x_4)F_{5/3}(\eta) - 
\left(\frac{1}{2}+x_4\right)F_{8/3}(\eta) \Biggr]n^{\beta+2}k_\mathrm{F}^2 \\
 + \frac{3}{40}t_5\Biggl[(2+x_5)F_{5/3}(\eta) + 
\left(\frac{1}{2}+x_5\right)F_{8/3}(\eta) \Biggr]n^{\gamma+2}\,k_\mathrm{F}^2 \quad , 
\end{multline}

\begin{equation}
\label{seqn2a}
k_\mathrm{F}= \left(\frac{3\pi^2n}{2}\right)^{1/3} \quad  ,
\end{equation}
\begin{equation}\label{seqn2b}
\eta = \frac{\rho_n - \rho_p}{\rho} = 1 - 2Y_p
\end{equation}
and
\begin{equation}
\label{seqn2c}
F_m(\eta)= \frac{1}{2}\Biggl[(1+\eta)^m+(1-\eta)^m\Biggr]\quad .
\end{equation}
Here $Y_p=Z/A=\rho_p/\rho$ is the proton fraction, $\rho_p$ and $\rho$ are the proton and baryonic number densities, respectively.

\begin{figure}[t]
\centering
\includegraphics[width=8cm]{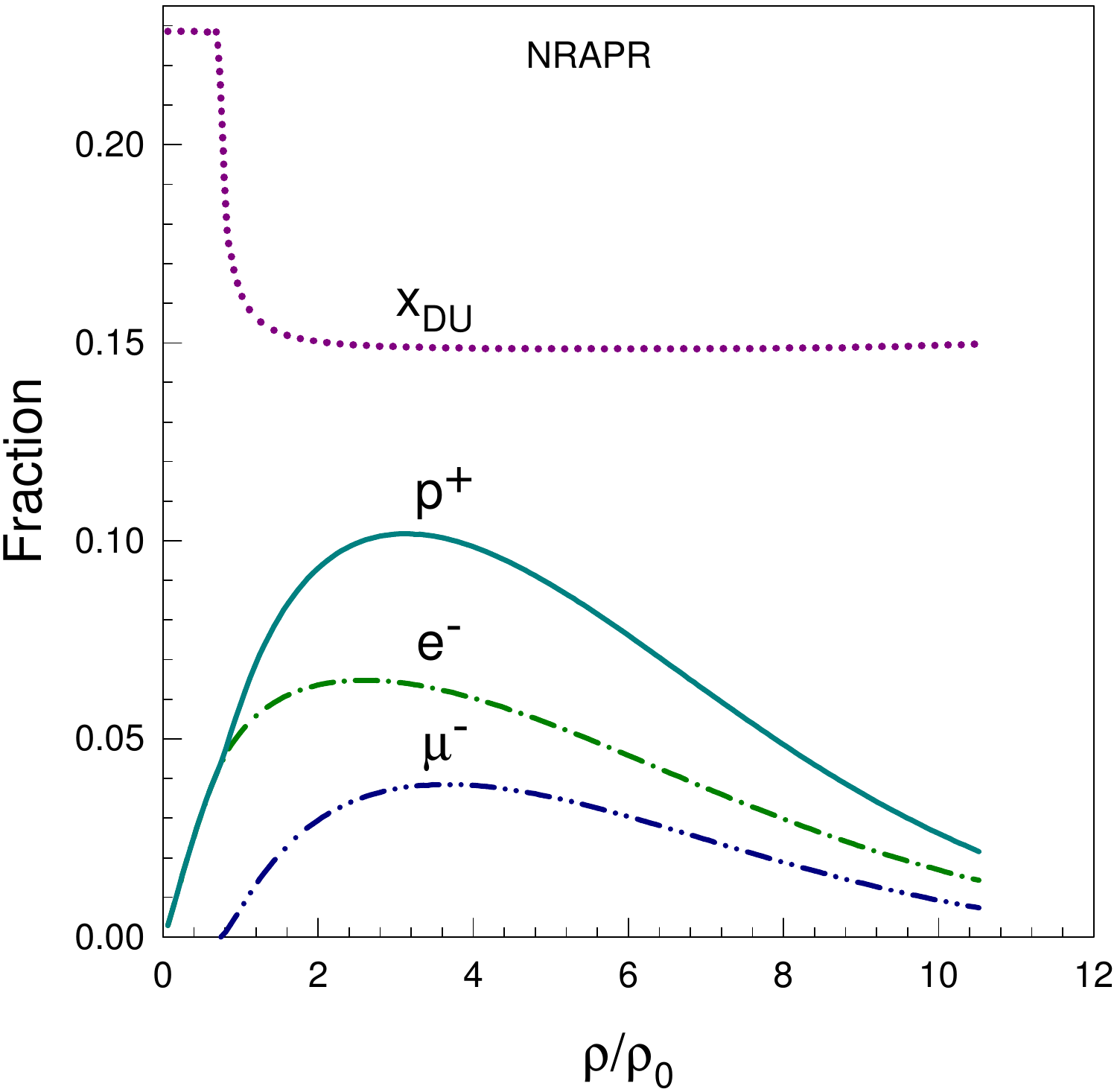}
\caption
{Plots of the $\beta$-equilibrium electron, muon and proton fractions as functions of the baryonic density for NRAPR interaction. The dURCA proton fraction threshold ${\mathrm x}_{\mathrm{DU}}$ is also shown.}
\label{fig1}
\end{figure}

    The pressure $P(\rho,\eta)$ in ANM can be defined in terms of energy per particle or chemical potential as 

\begin{eqnarray}
P(\rho,\eta)&=&\rho^2\frac{\partial\epsilon(\rho, Y_p)}{\partial\rho}\\ \nonumber
&=&\mu_n\rho_n + \mu_p\rho_p - H(\rho,\eta),\\ \nonumber
\label{seqn3}
\noindent
\end{eqnarray}
\noindent
where $H(\rho,\eta)$ represents energy density of ANM and is related to energy per particle $\epsilon(\rho,\eta)$ of ANM as $H(\rho,\eta)=\rho\epsilon(\rho,\eta)$ and
$\mu_{n(p)}~\Big(=\frac{\partial H(\rho,\eta)}{\partial\rho_{n(p)}}\Big)$ is the neutron(proton) chemical potential. The pressure $P(\rho,\eta)$ in ANM may then be expressed as

\begin{multline}\label{seqn4}
P(\rho,\eta)
=\frac{\hbar^2}{10}\left[\frac{1}{M_\mathrm{n}}(1+\eta)^{5/3}+
\frac{1}{M_\mathrm{p}}(1-\eta)^{5/3}\right]nk_\mathrm{F}^2    \\
+ \frac{1}{8}t_0\Biggl[3 - (1+2x_0)\eta^2\Biggr]n^2  \\
+ \frac{1}{8}t_1\Biggl[(2+x_1)F_{5/3}(\eta) -
\left(\frac{1}{2}+x_1\right)F_{8/3}(\eta) \Biggr]n^2k_\mathrm{F}^2 \\
 + \frac{1}{8}t_2\Biggl[(2+x_2)F_{5/3}(\eta) +
\left(\frac{1}{2}+x_2\right)F_{8/3}(\eta) \Biggr]n^2k_\mathrm{F}^2 \\
+ \frac{(\alpha + 1)}{48}t_3\Biggl[3-(1+2x_3)\eta^2\Biggr]n^{\alpha+2} \\
+ \frac{3\beta + 5}{40}t_4\Biggl[(2+x_4)F_{5/3}(\eta) -
\left(\frac{1}{2}+x_4\right)F_{8/3}(\eta) \Biggr]n^{\beta+2}k_\mathrm{F}^2 \\
+ \frac{3\gamma + 5}{40}t_5\Biggl[(2+x_5)F_{5/3}(\eta) + 
\left(\frac{1}{2}+x_5\right)F_{8/3}(\eta) \Biggr]n^{\gamma+2}\,k_\mathrm{F}^2 \quad. 
\\
\end{multline}

\begin{figure}[t]
\centering
\includegraphics[width=8cm]{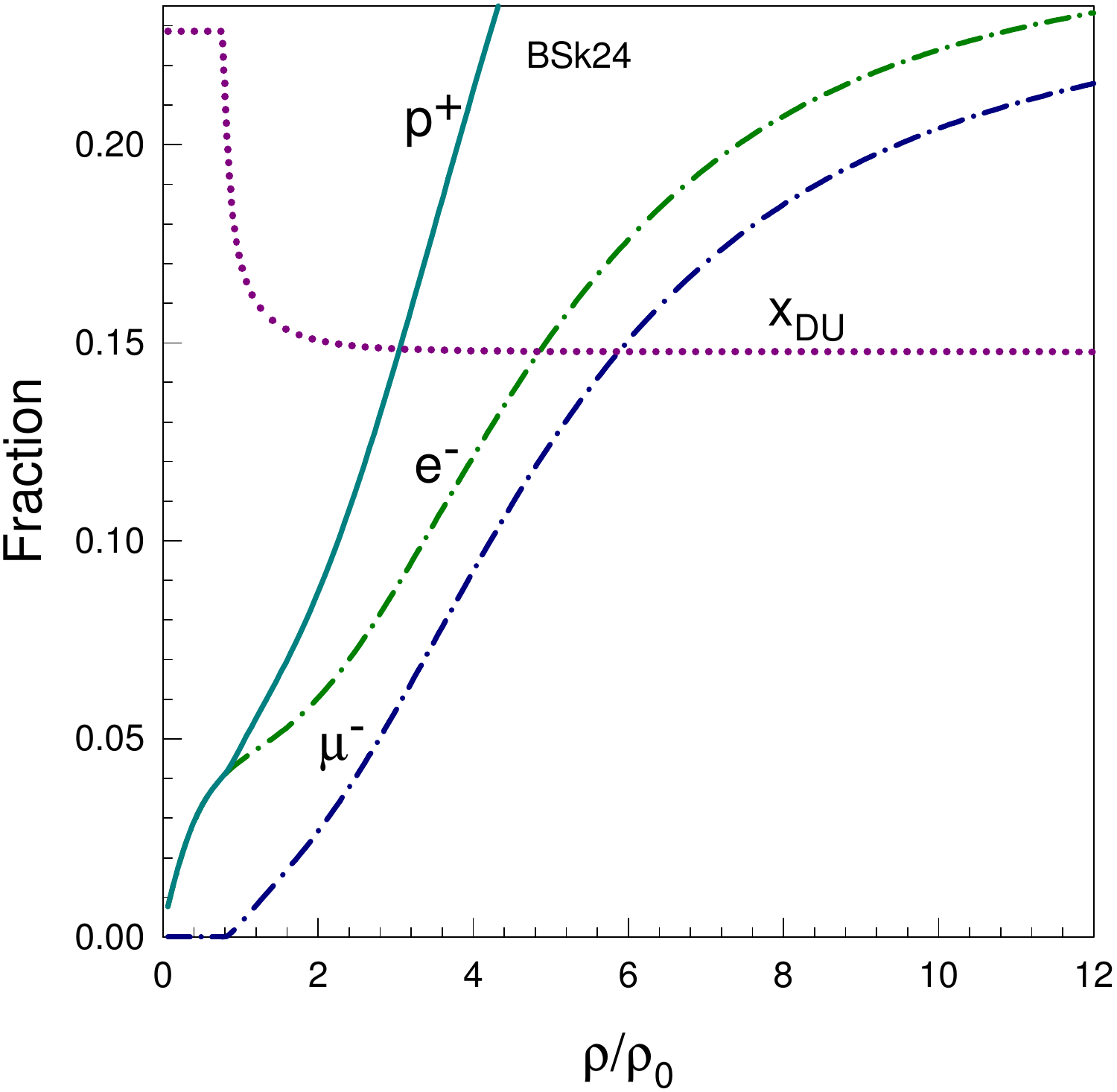}
\caption
{Same as Figure~\ref{fig1} but for BSk24 interaction.}
\label{fig2}
\end{figure}

\begin{figure}[t]
\centering
\includegraphics[width=8cm]{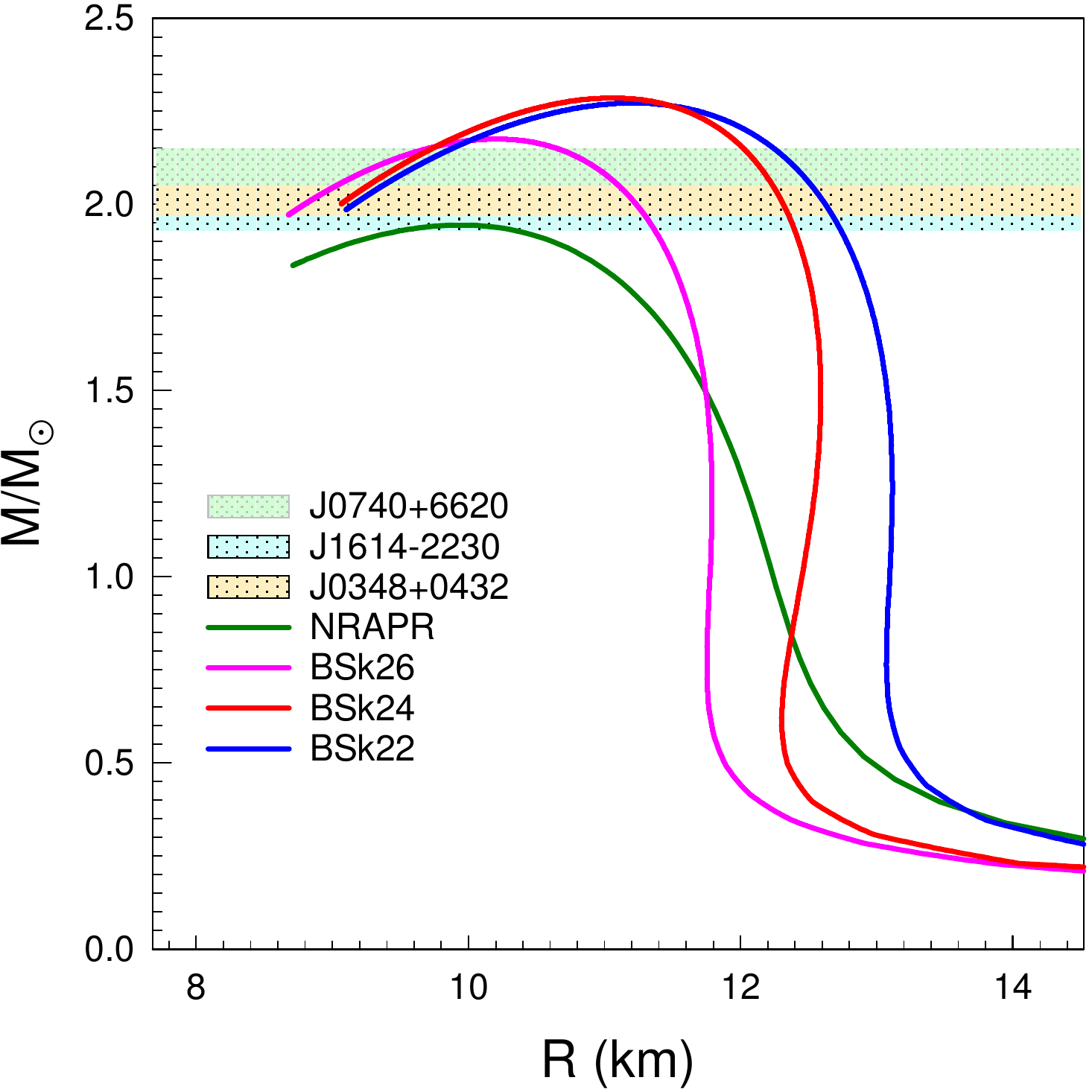}
\caption
{The mass-radius relation of slowly rotating neutron stars for the EoSs using NRAPR, BSk22, BSk24 and BSk26 interactions. Horizontal bands correspond to mass constraints of PSR J1614-2230, PSR J0348+0432 and PSR J0740+6620.}
\label{fig3}
\end{figure}
\noindent  

\section{The \texorpdfstring{$\beta$}{}-equilibrated cold dense \texorpdfstring{npe$\mu$}{} matter}

    The EoS has a relevancy on whether the dURCA process, where a NS cools very rapidly, can take place in NSs or not \cite{La91,Pe92}. Much attention has been drawn to dURCA process in NSs which may be the primary mechanism for its rapid cooling. This can, however, occur only when the $\beta$-equilibrium proton fraction $Y_p$ in the star is $\geq {\mathrm x}_{\mathrm{DU}}=1/9$, when only electrons are considered, and $\geq {\mathrm x}_{\mathrm{DU}}=1/[1+\{1+\frac{Y_e^{1/3}+(1-Y_e)^{1/3}}{2}\}^3]$, when both electrons and muons are considered where ${\mathrm x}_{\mathrm{DU}}$ is the dURCA proton fraction threshold. At very high densities or relativistic energies the leptonic masses can be neglected compared to their kinetic energies and then electronic number density $\rho_e$ $\approx$ muonic number density $\rho_\mu$ implying $Y_e=\rho_e/(\rho_e+\rho_\mu)\approx\frac{1}{2}$ and then $Y_p\geq {\mathrm x}_{\mathrm{DU}}= 1/[1+(1+2^{-1/3})^3] \approx~0.148$. In case of the dURCA process, the reactions which occur are the neutron decay $n \rightarrow p+e^-+\bar{\nu_e}$, electron capture $p+e^-\rightarrow n+ \nu_e$, neutron decay $n \rightarrow p+\mu^-+\bar{\nu_\mu}$ and muon capture $p+\mu^-\rightarrow n+ \nu_\mu$. In our study, the lepton energy per particle $\epsilon_L(\rho,Y_p)$ is given by the relativistic, ideal Fermi-gas expression. In addition to $e^-$, we also consider $\mu^-$ as and when they are energetically favored. At $\beta$-equilibrium, $\mu_n-\mu_p=\mu_L$ where $\mu_i$ is the chemical potential for $i_{th}$ species. As the chemical potential is the energy spent in creating a particle, $\mu_L=\frac{\partial}{\partial Y_p} \left (\epsilon_L(Y_p) \right )$, but $\mu_p-\mu_n=\frac{\partial}{\partial Y_p} \left ( \epsilon(\rho,Y_p) \right )$, since creation of a proton is associated with a neutron decay. This implies that $\frac{\partial}{\partial Y_p} \left ( \epsilon(\rho,Y_p)+\epsilon_L(Y_p) \right )=0$, where $\epsilon(\rho, Y_p)$ is the baryonic energy  per particle including the rest masses. 
 
\begin{figure}[t]
\centering
\includegraphics[width=8cm]{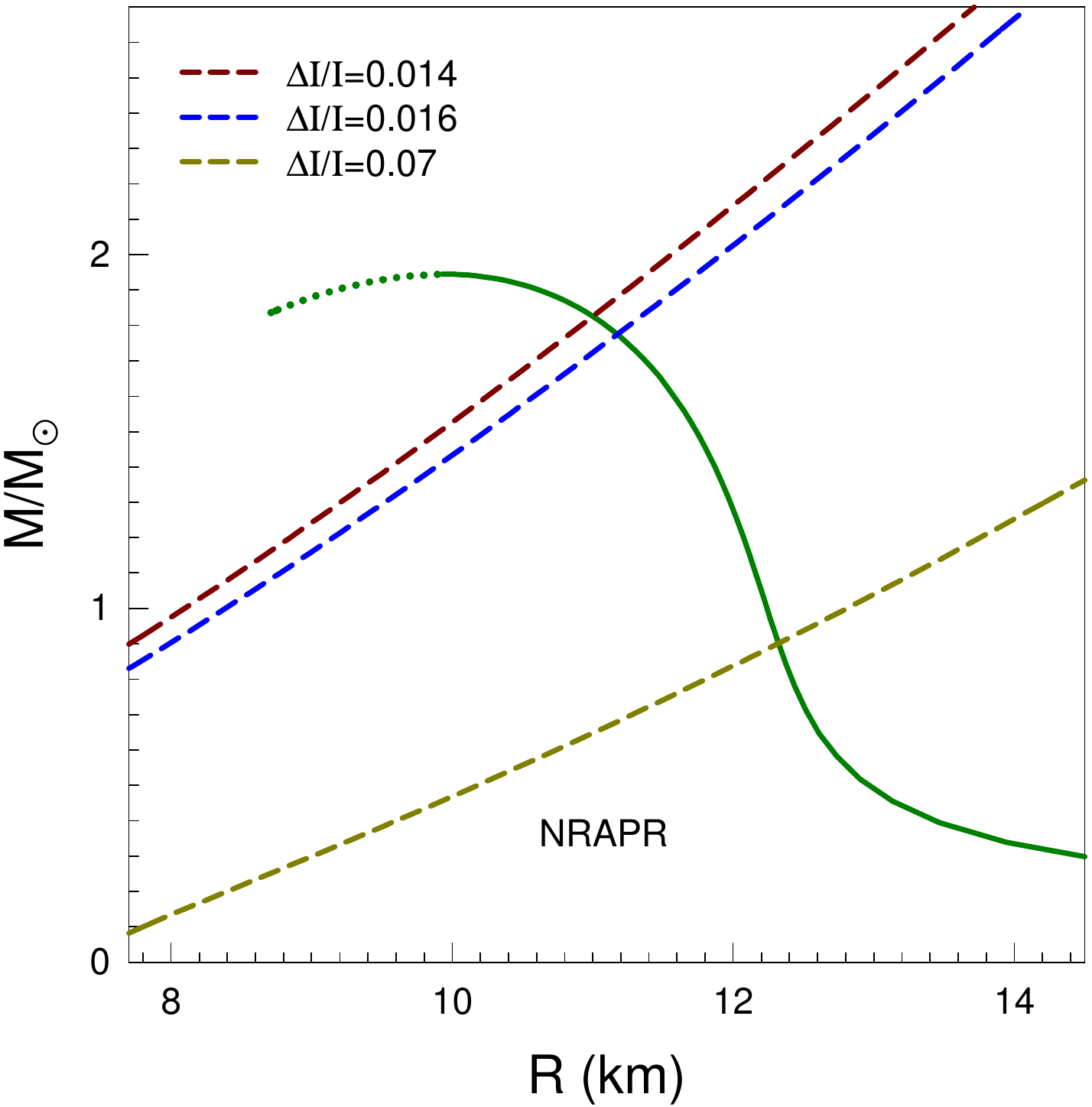}
\caption
{The mass-radius relation of slowly rotating neutron stars using the NRAPR EoS. For canonical Vela like pulsar, the constraint of $\Delta I/I > 1.4\%$ implies that allowed masses and radii lie to the right of the line defined by $\Delta I/I = 0.014$ (with $\rho_t=$ 0.073 fm$^{-3}$, P$_t=$ 0.413 MeV fm$^{-3}$). The unstable branch of TOV solution is shown in part by the dotted line.}
\label{fig4}
\end{figure}
\noindent 

\subsection{Tolman-Oppenheimer-Volkoff equation} 

    In general relativity, the structure of a spherically symmetric body of isotropic material which is in static gravitational equilibrium is given by the Tolman-Oppenheimer-Volkoff (TOV) equation \cite{TOV39a,TOV39b} 

\begin{eqnarray}
\frac{dP(r)}{dr} = -\frac{G}{c^4}\frac{[\varepsilon(r)+P(r)][m(r)c^2+4\pi r^3P(r)]}{r^2[1-\frac{2Gm(r)}{rc^2}]} \\ 
{\rm where} ~\varepsilon(r)=(\epsilon + m_b c^2)\rho(r),~m(r)c^2=\int_0^r \varepsilon(r') d^3r' \nonumber
\label{seqn5}
\end{eqnarray}
\noindent 
where $\varepsilon(r)$ and $P(r)$ are, respectively, the energy density and the pressure at a radial distance $r$ from the center of the star and $m(r)$ is the stellar mass contained within radius $r$. The numerical solution of TOV equation for masses and radii has been obtained using Runge-Kutta method. The EoS provides $\varepsilon(\rho)$ and $P(\rho)$ as the inputs for the calculations. The boundary condition $P(r)=0$ at the surface $R$ of the star determines its size and integration up to $R$ given by $M=m(R)$ \cite{Um97} provides the its total mass $M$. The numerical solution of TOV equation, being an initial value problem, requires a single integration constant ${\it viz.}$ the pressure $P_c$ at the center $r=0$ of the star calculated at a given central density $\rho_c$. It is important to mention here that the solution of TOV equation determines masses of static stars which are very close \cite{Ch10,Mi12,Ba14} to slowly rotating NSs.    

    The moment of inertia of a NS can be obtained by assuming it to be a slowly rotating object having a uniform angular velocity $\Omega$ \cite{Ha67,Ar77}. The angular velocity $\bar{\omega}(r)$ of a point in the star measured with respect to the angular velocity of the local inertial frame is governed by the equation

\vspace{-0.0cm}
\begin{equation}
\frac{1}{r^4}\frac{d}{dr}\left[r^4 j \frac{d\bar{\omega}}{dr}\right] + \frac{4}{r}\frac{dj}{dr}\bar{\omega}= 0
\label{seqn6}
\end{equation}
\noindent 
where

\vspace{-0.0cm}
\begin{equation}
j(r) = e^{-\phi(r)} \sqrt{1-\frac{2Gm(r)}{rc^2}}.
\label{seqn7}
\end{equation}
\noindent 
The function $\phi(r)$ is constrained by the condition

\vspace{-0.0cm}
\begin{equation}
e^{\phi(r)}\mu(r)={\rm constant}=\mu(R)\sqrt{1-\frac{2GM}{Rc^2}}
\label{seqn8}
\end{equation}
\noindent 
where the chemical potential $\mu(r)$ is defined as

\vspace{-0.0cm}
\begin{equation}
\mu(r) = \frac{\varepsilon(r)+P(r)}{\rho(r)}.
\label{seqn9}
\end{equation}
\noindent
Using these relations, Eq.~\ref{seqn6} can be solved subject to the boundary conditions that $\bar{\omega}(r)$ is regular as $r \rightarrow 0$ and $\bar{\omega}(r) \rightarrow \Omega$ as $r \rightarrow \infty$. The moment of inertia of the star can then be calculated using the definition $I = J/\Omega$, where the total angular momentum $J$ is given by

\vspace{-0.0cm}
\begin{equation}
J = \frac{c^2}{6G} R^4 \frac{d\bar{\omega}}{dr}\Big|_{r= R}.
\label{seqn10}
\end{equation}
\noindent

\begin{figure}[t]
\centering
\includegraphics[width=8cm]{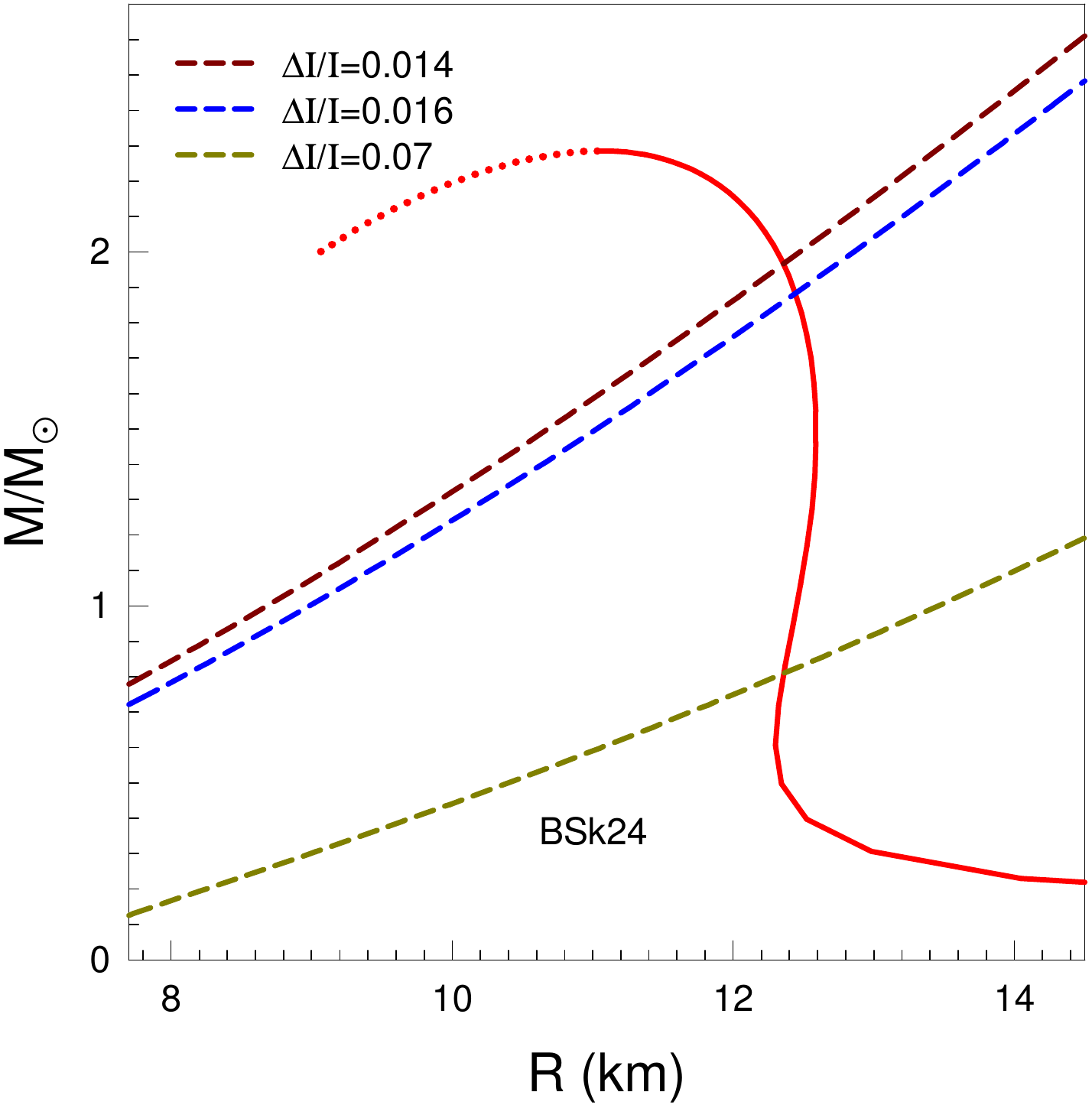}
\caption
{Same as Figure~\ref{fig4} but for BSk24 (with $\rho_t=$ 0.0807555 fm$^{-3}$, P$_t=$ 0.267902 MeV fm$^{-3}$).}
\label{fig5}
\end{figure}
\noindent 

\begin{table*}[htbp]
\centering
\caption{Values of the parameters of nuclear matter for the Skyrme interactions corresponding to NRAPR \cite{St05}, BSk22, BSk24 and BSk26 \cite{Gor13}. Since the parameters $t_2$ and $x_2$ always appear in the form of $t_2x_2$, the value of the latter is provided.}
\begin{tabular}{lcccc}
\hline
Parameter &NRAPR & BSk22& BSk24& BSk26 \\ \hline
$t_0$ [MeV $\mathrm{fm}^{3}$]          &-2719.7 &-3978.97  &-3970.29 &-4072.53  \\ 
$t_1$ [MeV $\mathrm{fm}^{5}$]          &417.64  &404.461   &395.766  &439.536  \\
$t_3$ [MeV $\mathrm{fm}^{3+3\alpha}$]  &15042	  &22704.7   &22648.6  &23369.1  \\
$t_4$ [MeV $\mathrm{fm}^{5+3\beta}$]   &	0.0   &-100.000  &-100.000 &-100.0  \\
$t_5$ [MeV $\mathrm{fm}^{5+3\gamma}$]  &	0.0   &-150.000  &-150.000 &-120.0  \\
$x_0$   &0.16154                &0.472558  &0.894371  &0.577367  \\
$x_1$   &-0.047986	            &0.0627540 &0.0563535 &-0.404961  \\
$t_2x_2$ [MeV $\mathrm{fm}^{5}$]&-1.811886&-1396.13&-1389.61  &-1147.70  \\
$x_3$   &0.13611 &0.514386  &1.05119  &0.624831  \\
$x_4$   & 0.0	   &2.00000   &2.00000  &-3.00000  \\
$x_5$   & 0.0	 	 &-11.0000  &-11.0000 &-11.00000  \\
$\alpha$& 0.0	 	 &1/12  &1/12 &1/12  \\
$\beta$ &	0.0	   &1/2   &1/2  &1/6  \\
$\gamma$&0.14416 &1/12  &1/12 &1/12  \\ \hline
\multicolumn{5}{c}{Nuclear matter properties at saturation density} \\ \hline
$\rho_0$ [$\mathrm{fm}^{-3}$]  &0.1606  &0.1578  &0.1578  &0.1589  \\ 
$\epsilon (\rho_0)$ [MeV]      &-15.86  &-16.088  &-16.048  &-16.064  \\
$K_\infty (\rho_0)$ [MeV]      &225.65	 &245.9  &245.5  &240.8  \\
$\frac{m^*}{m}(\rho_0,k_{f_0})$&0.69	 &0.80  &0.80  &0.80  \\
$E_{sym} (\rho_0)$ [MeV]       &32.79	 &32.0  &30.0  &30.0  \\ 
$L (\rho_0)$ [MeV]             &59.63	 &68.5  &46.4  &37.5  \\ \hline
\multicolumn{5}{c}{Parameters relating to the crust-core transition} \\ \hline
$\rho_t$ [$\mathrm{fm}^{-3}$]  &0.073 &0.0716068 &0.0807555 &0.0849477 \\
P$_t$ [MeV $\mathrm{fm}^{-3}$] &0.413 &0.290934 &0.267902 &0.363049 \\ 
\hline
\end{tabular}
\label{table1} 
\end{table*}

\begin{figure}[t]
\centering
\includegraphics[width=8cm]{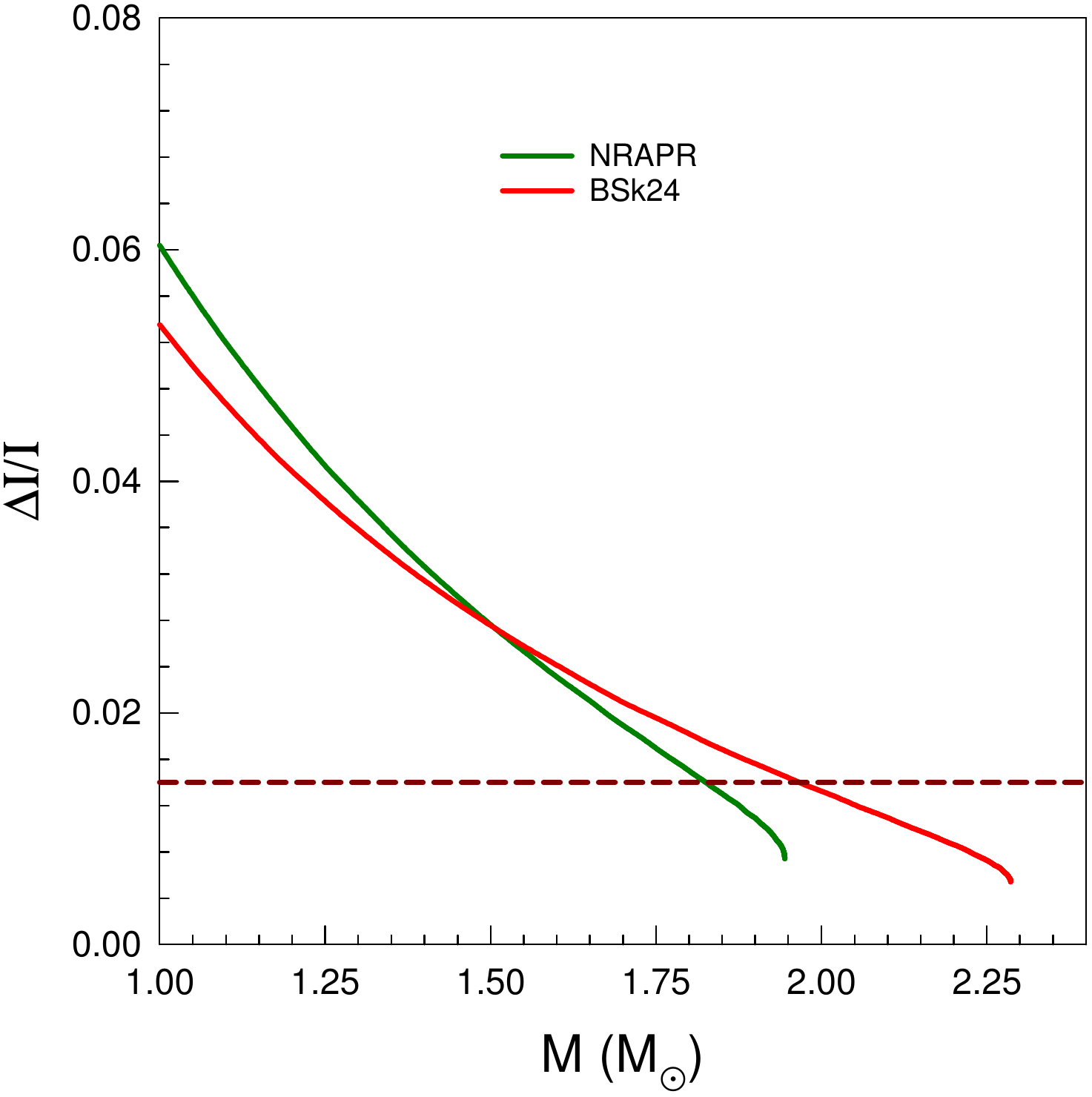}
\caption
{Plots of crustal fraction of moment of inertia versus NS mass for NRAPR and BSk24 EoSs. The dashed line defined by $\Delta I/I = 0.014$ represents the lowest limit of $\Delta I/I$ for canonical Vela like pulsar.}
\label{fig6}
\end{figure}

\noindent
\section{Results and discussion}
\label{section3}

    The pulsars, which are NSs with masses $>$ 1.8 $M_\odot$, are critical probes of nuclear astrophysics in extreme conditions. These massive NSs have extremely high gravitational fields inside, leading to substantially higher gravitational binding energies, than those commonly found in canonical 1.4 $M_\odot$ NSs. In the present work, the EoSs for $\beta$-equilibrated NS matter have been studied for the Skyrme interactions with NRAPR \cite{St05}, BSk22, BSk24 and BSk26 \cite{Gor13} parameter sets. These parameters along with the values of SNM incompressibility $K_\infty$, symmetry energy $E_{sym}(\rho_0)$ and slope of nuclear symmetry energy $L$ are provided in Table~\ref{table1}. The core-crust transition density $\rho_t$ and pressure P$_t$ determined by the dynamical method have also been tabulated. 

\subsection{The proton fraction and the URCA process}
    
    In Figures~\ref{fig1} and \ref{fig2}, the $\beta$-equilibrium electron fraction, muon fraction and proton fraction have been plotted as a function of $\rho/\rho_0$ for NRAPR and BSk24 respectively. Since both electrons and muons have been considered in the calculations and the resultant $\beta$-equilibrium proton fraction is always less than the dURCA threshold proton fraction ${\rm x}_{\rm DU}$ \cite{La91}, the calculated value of $x_p$ forbids dURCA process for NRAPR but permits for BSk22, BSk24 and BSk26. An EoS describing ANM may not allow the occurrence of dURCA process in NSs with masses $<$ 1.5 $M_\odot$ theoretically \cite{AWS06,Ca06,Kl06}, but for an EoS leading to higher NS masses this process is possible. It may be recalled that the neutrino emission in NS is caused due to the processes of dURCA, modified URCA, Cooper-pair creation and breaking \cite{Cr10,Da11,Dm11,Sh11} and nucleon-nucleon bremsstrahlung. The dURCA process is the fastest cooling mechanism for which the neutrino emissivity can be about six orders of magnitude larger than other mechanisms. The dURCA process is believed to occur early in the lives of all NSs and may operate in later stages of a few massive NSs \cite{La18}.			
		
\begin{table*}[htbp]
\centering
\caption{Radius, mass, crustal thickness, crustal mass and crustal fraction of moment of inertia as functions of central density $\rho_c$ for NRAPR (for $\rho_t=$ 0.073 $\mathrm{fm}^{-3}$, P$_t=$ 0.413 MeV $\mathrm{fm}^{-3}$).}
\begin{tabular}{c|c|c|c|c|c|c}
\hline
$\rho_c$ [fm$^{-3}$]&$R$ [km]&$M$ [$M_\odot$]&$\Delta R$ [km]&$\Delta M$ [$M_\odot$]&$I$ [$M_\odot$km$^2$]&$\Delta I/I$\\ \hline

 1.30&   9.8825&   1.9442&    .3301&    .0056&  97.85       &  .0073 \\
 1.25&   9.9943&   1.9442&    .3425&    .0059&  99.85       &  .0077 \\
 1.20&  10.1105&   1.9420&    .3566&    .0063& 101.76       &  .0082 \\
 1.15&  10.2310&   1.9370&    .3724&    .0067& 103.56       &  .0087 \\
 1.10&  10.3560&   1.9288&    .3907&    .0071& 105.20       &  .0093 \\
 1.05&  10.4851&   1.9168&    .4114&    .0077& 106.63       &  .0100 \\
 1.00&  10.6185&   1.9003&    .4353&    .0083& 107.76       &  .0109 \\
  .95&  10.7557&   1.8785&    .4631&    .0090& 108.51       &  .0118 \\
  .90&  10.8963&   1.8504&    .4954&    .0098& 108.79       &  .0130 \\
  .85&  11.0396&   1.8151&    .5334&    .0108& 108.48       &  .0144 \\
  .80&  11.1850&   1.7714&    .5785&    .0119& 107.44       &  .0161 \\
  .75&  11.3312&   1.7179&    .6325&    .0133& 105.55       &  .0182 \\
  .70&  11.4768&   1.6534&    .6978&    .0148& 102.54       &  .0209 \\
  .65&  11.6203&   1.5763&    .7779&    .0167&  98.36       &  .0241 \\
  .60&  11.7598&   1.4855&    .8773&    .0189&  92.89       &  .0283 \\
  .55&  11.8933&   1.3798&   1.0028&    .0216&  86.01       &  .0338 \\
  .50&  12.0199&   1.2586&   1.1644&    .0249&  77.70       &  .0408 \\
  .45&  12.1408&   1.1223&   1.3778&    .0288&  68.10       &  .0503 \\
  .40&  12.2624&    .9723&   1.6683&    .0336&  57.51       &  .0629 \\
  .35&  12.4035&    .8121&   2.0794&    .0394&  46.46       &  .0801 \\
  .30&  12.6119&    .6473&   2.6931&    .0465&  35.73       &  .1037 \\
  .25&  13.0102&    .4860&   3.6813&    .0553&  26.28       &  .1374 \\
  .20&  13.9441&    .3386&   5.4708&    .0658&  19.23       &  .1917 \\
\hline
\end{tabular}
\label{table2} 
\end{table*}
\noindent 		

\begin{table*}[htbp]
\centering
\caption{Radius, mass, crustal thickness, crustal mass and crustal fraction of moment of inertia as functions of central density $\rho_c$ for BSk24 (for $\rho_t=$ 0.0807555 $\mathrm{fm}^{-3}$, P$_t=$ 0.267902 MeV $\mathrm{fm}^{-3}$).}
\begin{tabular}{c|c|c|c|c|c|c}
\hline
$\rho_c$ [fm$^{-3}$]&$R$ [km]&$M$ [$M_\odot$]&$\Delta R$ [km]&$\Delta M$ [$M_\odot$]&$I$ [$M_\odot$km$^2$]&$\Delta I/I$\\ \hline

 1.00&  10.9911&   2.2853&    .2908&    .0044& 152.82        &  .0053 \\
  .95&  11.1456&   2.2850&    .3060&    .0048& 156.93        &  .0057 \\
  .90&  11.3043&   2.2793&    .3231&    .0052& 160.74        &  .0061 \\
  .85&  11.4672&   2.2666&    .3428&    .0056& 164.03        &  .0067 \\
  .80&  11.6353&   2.2453&    .3681&    .0062& 166.51        &  .0074 \\
  .75&  11.8040&   2.2131&    .3979&    .0069& 167.90        &  .0083 \\
  .70&  11.9723&   2.1672&    .4351&    .0077& 167.73        &  .0094 \\
  .65&  12.1342&   2.1041&    .4805&    .0086& 165.51        &  .0108 \\
  .60&  12.2865&   2.0197&    .5390&    .0098& 160.59        &  .0128 \\
  .55&  12.4202&   1.9092&    .6153&    .0113& 152.26        &  .0154 \\
  .50&  12.5236&   1.7674&    .7162&    .0131& 139.86        &  .0191 \\
  .45&  12.5811&   1.5892&    .8532&    .0154& 122.99        &  .0244 \\
  .40&  12.5797&   1.3715&   1.0521&    .0182& 101.62        &  .0326 \\
  .35&  12.5032&   1.1161&   1.3510&    .0217&  76.93        &  .0457 \\
  .30&  12.3700&    .8341&   1.8409&    .0259&  51.58        &  .0675 \\
  .25&  12.3096&    .5511&   2.7412&    .0309&  29.73        &  .1048 \\
  .20&  12.9826&    .3072&   4.8084&    .0368&  15.65        &  .1755 \\
\hline
\end{tabular}
\label{table3} 
\end{table*}
\noindent

\subsection{The mass-radius relationship for neutron stars}

    The calculations for masses and radii have been performed employing the EoSs FMT \cite{FMT}, BPS \cite{Ba71} and BBP \cite{Ba71a} up to the number density of 0.0582 fm$^{-3}$ for the outer crust and the $\beta$-equilibrated NS matter beyond covering the inner crustal and core regions of a compact star. The location of the inner edge of the NS crust, the core-crust transition density and pressure can be determined by dynamical method, thermodynamical method and RPA. It is important to mention here that both the dynamical and the thermodynamical methods give very similar results with the former giving a slightly smaller transition density than the latter and this is due to the fact that the former includes the density gradient and Coulomb terms that make the system more stable and lower the transition density. The small difference between the two methods implies that the effects of density gradient terms and the Coulomb term are unimportant in determining the transition density \cite{Xu09}. In the present calculations the dynamical method has been used. As may be seen from Table~\ref{table1}, the number density at the core-crust transition for NRAPR \cite{Du11}, BSk22, BSk24 and BSk26 \cite{Pea18}, is far beyond the outer crustal part covered by FMT+BPS+BBP, implying that the core-crust transition between the core and the inner crust has been treated consistently within the same theoretical model.
    
    In Figure~\ref{fig3}, the mass-radius relationship is plotted for slowly rotating NSs for EoSs obtained with NRAPR, BSk22, BSk24 and BSk26 by solving the TOV equations. The maximum NS mass for the EoS obtained using the NRAPR Skyrme set is 1.94 $M_\odot$ with a corresponding radius of 9.93 kms while BSk22, BSk24 and BSk26 interactions yield masses 2.27 $M_\odot$, 2.28 $M_\odot$ and 2.18 $M_\odot$ with corresponding radii of 11.21 kms, 11.05 kms and 10.21 kms, respectively. 
		
    It is worthwhile to mention that the observations of the binary millisecond pulsar J1614-2230 by Demorest et al.~\cite{De10} suggest that the masses lie within 1.97 $\pm$ 0.04 $M_\odot$. The radio timing measurements of the pulsar PSR J0348+0432 and its white dwarf companion have confirmed the mass of the pulsar to be in the range 2.01 $\pm$ 0.04 $M_\odot$ \cite{An13}. Very recently the observations for PSR J0740+6620 \cite{Rez18,Fon21} and for PSR J0952-0607 \cite{Rom22} find masses of 2.08 $\pm$ 0.07 $M_\odot$ and 2.35 $\pm$ 0.17 $M_\odot$, respectively. Some recent works \cite{Leg21,Ril21} constrain the mass and equatorial radius of PSR J0740+6620 to be 2.072$^{+0.067}_{-0.066}$ $M_\odot$ and 12.39$^{+1.30}_{-0.98}$ km respectively. Moreover, the recent observations of the gravitational wave event GW170817 estimate the maximum mass as ${2.01}_{-0.04}^{+0.04}\leqslant {M}_{\mathrm{TOV}}/{M}_{\odot }\lesssim {2.16}_{-0.15}^{+0.17}$ \cite{Rez18}. The horizontal bands in Figure~\ref{fig3} correspond to mass constraints of PSR J1614-2230, PSR J0348+0432 and PSR J0740+6620. 		 		

\subsection{Core-crust transition and crustal fraction of moment of inertia}

    The crustal fraction of the moment of inertia $\Delta I/I$ can be expressed as a function of the gravitational mass $M$ and the radius $R$ of the star depending on the EoS arising from the values of transition density $\rho_t$ and pressure P$_t$ \cite{Li99}

\begin{eqnarray}
\frac{\Delta I}{I} \approx \frac{28\pi {\rm P}_t R^3}{3Mc^2} \left(\frac{1-1.67\xi-0.6\xi^2}{\xi}\right) \nonumber\\
\times \left(1+\frac{2{\rm P}_t}{\rho_t m_b c^2}\frac{(1+7\xi)(1-2\xi)}{\xi^2} \right)^{-1}
\label{seqn11}
\end{eqnarray} 
where $\xi=\frac{GM}{Rc^2}$. The crustal fraction of the moment of inertia is of interest since it can be extracted from the observed glitches of NSs. It has been shown \cite{Li99} that the glitches represent a self-regulating instability during which the star prepares over a waiting time. The angular momentum requirements of glitches in pulsars suggest that over $1.4\%$ of the total moment of inertia is involved in these events \cite{Li99}. Therefore, if glitches originate in the liquid of the inner crust, this implies that $\Delta I/I > 0.014$.

    The values of $I$ obtained by solving Eq.~\ref{seqn10} subject to the boundary conditions stated earlier are listed in Tables~\ref{table2} and \ref{table3} along with mass $M$, radius $R$, crustal thickness $\Delta R$ and crustal mass $\Delta M$ of NSs for NRAPR and BSk24 respectively. Once masses and radii have been determined, the $\Delta I/I$ are obtained from Eq.~\ref{seqn11} and are also tabulated there. From Eq.~\ref{seqn11} the mass-radius relation has been extracted as well for fixed values of $\Delta I/I$, $\rho_t$ and P$_t$. In Figure~\ref{fig4}, this mass versus radius is plotted for $\Delta I/I = 0.014$ for NRAPR, while Figure~\ref{fig5} depicts the same for BSk24. The core-crust transition density and pressure (provided in Table~\ref{table1}) together with the extracted minimum crustal fraction of the total moment of inertia $\Delta I/I>1.4\%$ provide a limit for the radii of pulsars which may be elegantly expressed in the form of inequalities $R \geq 3.71 + 3.78 M/M_\odot$ km for NRAPR, $R \geq 4.03 + 3.83 M/M_\odot$ km for BSk22, $R \geq 3.90 + 3.91 M/M_\odot$ km for BSk24 and $R \geq 3.67 + 3.81 M/M_\odot$ km for BSk26.
    
    In Figure~\ref{fig6} the crustal fraction of moment of inertia has been plotted with NS mass for NRAPR and BSk24 EoSs. The dashed line defined by $\Delta I/I = 0.014$ represents the lowest limit of $\Delta I/I$ for canonical Vela like pulsar. Our results imply that the crust may be sufficient \cite{Pi14} for pulsars with masses $\leq$ 1.82 $M_\odot$ in case of NRAPR and $\leq$ 1.97 $M_\odot$ for BSk24 with a crustal fraction of the total moment of inertia $\Delta I/I>1.4\%$. It is suggested \cite{Ho12} that this value may at most be 1.6$\%$, which does not cause any significant difference. This marginally higher value of $\Delta I/I$ does not extend the partial superfluidity into the core. The analysis of about 1 $M_\odot$ canonical Vela-like pulsar glitches requires the crustal fraction of the moment of inertia to exceed 1.4$\%$ if entrainment is ignored as concluded in Ref.~\cite{Li99}. However, if entrainment is considered, the lower bound is raised to about 7$\%$ \cite{An12,Ch13}. In those studies nondissipative crustal entrainment, wherein the neutron superfluid is strongly coupled to the crust, caused due to the Bragg reflection of an unbound neutron by the lattice ion was accounted for. From Tables~\ref{table2}, \ref{table3} and Figure~\ref{fig6}, this would mean that the crustal fraction of the total moment of inertia is $\sim$6.0$\%$ (NRAPR), $\sim$6.3$\%$ (BSk22), $\sim$5.4$\%$ (BSk24) and $\sim$5.3$\%$ (BSk26) due to crustal entrainment, which is close to the lower bound provided in Refs.\cite{An12,Ch13}.
     
\noindent
\section{ Summary and conclusion }
\label{section4}

    The mass-radius relation of NSs is obtained by using EoSs from NRAPR, BSk22, BSk24 and BSk26 interactions. The structural properties of crust have also been studied using these EoSs. The calculations of the crustal mass, crustal radius, crustal fraction of moment of inertia require the transition density and pressure at core to crust transition apart from the bulk properties of the NSs. The maximum NS masses attained in these calculations are in conformity with recent observations, apart from several other nuclear matter constraints. The NS core-crust transition density and pressure determined using the dynamical method. These values along with the observed minimum crustal fraction of the total moment of inertia impose constraint on the mass-radius relation of NSs. The implications of these EoSs on crustal entrainment have also been explored.
   
\begin{acknowledgements}

    One of the authors (DNB) acknowledges support from Science and Engineering Research Board, Department of Science and Technology, Government of India, through Grant No. CRG/2021/007333.	
					
\end{acknowledgements}

\end{document}